\documentclass[11pt,leqno]{article}

\date{}  

\usepackage{url}
\usepackage{cite}
\usepackage{plain}

\usepackage{wrapfig}
\usepackage{graphics}
\usepackage{picins,graphicx}
\usepackage[english]{babel}
\usepackage[leqno]{amsmath}
\usepackage{amssymb}
\usepackage{verbatim}
\usepackage[mathscr]{eucal}

\usepackage{amstext}
\usepackage{makeidx}
\usepackage{amsthm}

\usepackage{hyperref}
\hypersetup{colorlinks,%
citecolor=black,%
filecolor=black,%
linkcolor=black,%
urlcolor=black}

\makeindex

\newtheorem{theorem}{Teorema} 

\newtheorem{proposition}{Proprieta'}
\newtheorem{definition}{Definizione}
\newtheorem{notation}{Nota}
\newtheorem{ex}{Esercizio} 
\newtheorem{esempio}{Esempio}

\newcommand{\vs}{\vspace{3mm}}
 
\newcommand{\no}{\noindent} 

\newcommand{\beq}{\begin{equation}} 
\newcommand{\eeq}{\end{equation}}

\newcommand{\bex}{\begin{ex}} 
\newcommand{\eex}{\end{ex}} 

\newcommand{\bese}{\begin{esempio}} 
\newcommand{\eese}{\end{esempio}} 

\newcommand{\bpro}{\begin{proposition}} 
\newcommand{\epro}{\end{proposition}} 

\newcommand{\ds}{\displaystyle}

\newcommand{\bthe}{\begin{theorem}} 
\newcommand{\ethe}{\end{theorem}}

\newcommand{\bnote}{\begin{notation}} 
\newcommand{\enote}{\end{notation}}

\newcommand{\bdefi}{\begin{definition}} 
\newcommand{\edefi}{\end{definition}} 

\newcommand{\bc}{\begin{center}} 
\newcommand{\ec}{\end{center}}

\newcommand{\mail}[1]{\href{unina:#1}{\texttt{#1}}}

\usepackage{pdfsync}

\author{Monica De Angelis\thanks{Univ. of Naples  "Federico II", Faculty of Engineering, Dip. Mat. Appl. "R.Caccioppoli", \newline
 Via Claudio n.21, 80125, Naples, Italy.
\newline\mail{modeange@unina.it}}}

\title{On a Model of Superconductivity and Biology}

\begin{document}
\maketitle

\begin{abstract}
\vs \no The paper deals with a semilinear integrodifferential equation that  characterizes several dissipative  models   of Viscoelasticity, Biology and Superconductivity. The initial - boundary problem with  Neumann  conditions is analyzed. When the source term $ F  $  is a linear function, then the explicit solution is obtained. When $ F $   is non linear, some results on existence, uniqueness  and a priori estimates are deduced. As example of physical model the reaction -  diffusion system of Fitzhugh Nagumo is considered.

\vs \no {\bf{Keywords}}:{ Reaction - diffusion systems;\hspace{2mm} Biological applications;\hspace{2mm} Laplace  transform }

\vs \no \textbf{Mathematics Subject Classification (2000)}\hspace{1mm}35E05  \hspace{1mm}35K35\hspace{1mm} 35K57 \hspace{1mm}35Q53 \hspace{1mm} 78A70
\end{abstract}


\vspace{3mm}


\section{Introduction }

\setcounter{equation}{0}
\setcounter{figure}{0}

\setcounter{definition}{0}
\setcounter{notation}{0}
\setcounter{theorem}{0}
\setcounter{esempio}{0}
\setcounter{ex}{0}
\setcounter{proposition}{0}
\setcounter{criterio}{0}

Let  consider a function  $ \,  u= u(x,t), \, $  where  $\, x \, $ is a direction of propagation and $\,t \, $ is the time,  and let

  \beq            \label{11}
     {\cal L}\,u\, \equiv \,  u_t -  \varepsilon  u_{xx} + au +b \int^t_0  e^{- \beta (t-\tau)}\, u(x,\tau) \, d\tau = \, F(\,x,\,t,\,u \,(x,t)\, ) \,    
 \eeq

\vs \no with $\,  a, \,\, b,\,\,\, \varepsilon, \, \beta \,  $    positive constants. 

\vs The  equation  (\ref{11})  describes  the evolution of several  physical models  as motions of viscoelastic fluids or solids  \cite{bcf,dr1,r}; heat conduction at low temperature \cite{mps}, sound propagation in viscous gases \cite{l}. Other two specifical examples for the integro differential equation  (\ref{11}) are related to  biological models and superconductivity.

\vs As for  the biological phenomena, a  well known reaction diffusion model is given by  the Fitzhugh - Nagumo system (FHN) \cite{i,m1,acscott02}: 

\beq     \label{12}
  \left \{
   \begin{array}{lll}
    \displaystyle{\frac{\partial \,u }{\partial \,t }} =\,  \varepsilon \,\frac{\partial^2 \,u }{\partial \,x^2 }
     \,-\, v\,\,  + f(u ) \,  \\
\\
\displaystyle{\frac{\partial \,v }{\partial \,t } }\, = \, b\, u\,
- \beta\, v\,.
\\

   \end{array}
  \right.
 \eeq

\vs\vs  \noindent In this case the function $\,f (u)\,$ is:

\beq      \label{13}
f(u)\, =\,-\,a\,u\, +\,\varphi(u) \quad with \quad  \varphi \,=\, u^2\, (\,a+1\,-u\,)   \qquad  \,(0<a<1)\, 
\eeq

\vs
As for the variables $\,u \,$ and $\,v \,$,  $\, u(x,t)\,$  represents  a membrane  potential  of  a  nerve  axon  at  distance  x and  time t, and  $\,v(x,t)\,$  is a  recovery variable that  models the  transmembrane current.

\vs   If $\,v_0  \,$ represents the initial value of v, the system (\ref{12}) can be given the form (\ref{11}) with 

\beq     \label{14}
F(x,t,u)\, =\,\varphi (u) \, -\, v_0(x) \, e^{\,-\,\beta\,t\,}.
\eeq

\vs Moreover, equation (\ref{11})  occurs also in superconductivity to describe the  Josephson tunnel effects in junctions. In this case the unknown  $\, u\, $ denotes the difference between the phases of the wave functions of the two superconductors and the differential equation is:

\beq  \label{15}
\varepsilon u_{xxt}\, - \, u_{tt} \, +\, u_{xx}-\, \alpha u_t = \,  \, \sin u \ + \gamma   
\eeq 

\vs \no where $\, \gamma \, $ is a constant  forcing term that is proportional to a bias current. The  $ \varepsilon -term$  and the $\, \alpha -term $ account for  the dissipative normal electron current  flow along and across the junction, respectively \cite{bp,acscott}.

From (\ref{11}) one obtains  the equation (\ref{15})  as soon as one assumes

\beq   \label{16}
 a \,=\,  \alpha \, - \dfrac{1}{\varepsilon} \, \,\quad\,\, b = \,  - \, \dfrac{a}{\varepsilon}  \,\,\quad \displaystyle \, \beta \,= \dfrac{1}{\varepsilon}\,\,
  \eeq
  
  \no  and $ F  $ is such that

\beq   \label{17}
F(x,t,u)\,=\, -\, \int _0^t \, e^{\,-\,\frac{1}{\varepsilon}\,(t-\tau\,)}\,\,[\, \, \gamma\,+\, sen \, u (x, \tau)\, \,]\, \, d\tau. 
\eeq

\vs As (\ref{16}) show, in the superconductive case the constants $ \,a,\,  b\,$ could be negative too.

\vs  The explicit fundamental solution $ K_0(x,t)  $ of the operator  $ \,{\cal L}\, $  defined in (\ref{11}) has been  already determined in  \cite{dr8}   together with numerous basic properties.  When $ F $ is a linear function, by means of  $ K_0(x,t)  $  it is possible to obtain the explicit solution of  both the    Neumann and  the Dirichelet   problem  for  (\ref{11}). When $\, \,F \,$ is non linear, an appropriate analysis of the  integro differential equation implies results on the existence, uniqueness and a priori estimates of the solution. These results will be applied to (FHN) system.

\section{Statement of the problem and tranform solution  }
\setcounter{equation}{0}

If  $\, T\, $  is  an arbitrary positive constant and 

\[
\,   \Omega_T \, \equiv \{\,(x,t) : \, 0\,\leq \,x \,\leq L \,\,;  \ 0 < t \leq T \, \},
\]

\vs\no let ($ P_{N} $) the  following Neumann initial - boundary value  problem  related to equation (\ref{11}):

\beq   \label{21}
\left \{
   \begin{array}{lll}
   u_t -  \varepsilon  u_{xx} + au +b \int^t_0  e^{- \beta (t-\tau)}\, u(x,\tau) \, d\tau \,=\, F(x,t,u) \, & (x,t) \in \Omega_T \,  \\    
\\
  \,u (x,0)\, = u_0(x)\, \,\,\, &
x\, \in [0,L], 
\\
 \\
  \, u_x(0,t)\,=\,\psi_1(t)  \qquad u_x(L,t)\,=\,\psi_2(t)  & 0<t\leq T.
   \end{array}
  \right.
\eeq

\vs\vs  \no In excitable systems this problem occurs when two-species reaction diffusion system is subjected to flux boundary condition \cite{m2}. The same conditions  are present in case of pacemakers \cite{ks}.  Neumann  conditions are applied also to study distributed FHN system \cite{ns}. 

 \no In superconductivity, instead, $( P_{N})$ problem  can be referred to the boundary specification  of the magnetic field \cite{bcs0,j,sfgpcs}.

\vs When   $ F\,= f(x,t) $ is a linear function,the problem $ (P_{N}) $  can  be solved by Laplace transform with respect to $ t. $

\newpage
  If

 \beq   \label{22}
 \left \{
   \begin{array}{lll} 
\hat u (x,s) \, = \int_ 0^\infty \, e^{-st} \, u(x,t) \,dt \,& \hat f (x,s)   \, = \int_ 0^\infty \,\, e^{-st} \,\, f\,(x,t) \,dt \,,
\\ 
\\
\hat \psi_i \,\,(s) \, = \int_ 0^\infty \, e^{-st} \,\, \psi_i\,(t)\, \,dt \,\,&(i=1,2),
 \end{array}
  \right.
\eeq

\vs\no one deduces the following transform   $ ( \hat P_N) $ problem:

\beq   \label{23}
\left \{
   \begin{array}{lll}
  \hat u_{xx}  \,\,- \frac{\sigma^2}{\varepsilon} \,\,\hat u =\, -\,\frac{1}{\varepsilon} \,\,[\, \,\hat f(x,s) +u_0(x)\,\,]\\    
\\
  \,\hat u_x(0,s)\,=\, \hat \psi_1\,(s)\qquad \hat u_x(L,s)\,=\,\hat \psi_2\,(s),
   \end{array}
  \right.
\eeq

\vs  \no    where $ \,\,\,\ds \sigma^2 \ \,=\, s\, +\, a \, + \, \frac{b}{s+\beta}.\,\,\,$  Letting  $\displaystyle\,\tilde{\sigma}^2\,=\, \sigma^2/{\varepsilon,}\,\, $ and considering the following function

\beq\,  \label{24}
\displaystyle
\hat \theta_0 \,(\,y,\tilde\sigma)\,= \,\dfrac{\cosh\,[\, \tilde\sigma \,\,(L-y)\,]}{\,2\, \, {\varepsilon} \,\,\tilde \sigma\,\,\, \sinh\, (\,\tilde \sigma \,L\,)}\,\,=
\eeq

\[ =\,  \frac{1}{2 \,\, \sqrt\varepsilon \,\,\,\sigma  } \, \biggl\{\, e^{- \frac{y}{\sqrt \varepsilon} \,\,\sigma}+\, \sum_{n=1}^\infty \,\, \biggl[ \,e^{- \frac{2nL+y}{\sqrt \varepsilon} \,\,\sigma} \, +\, e^{- \frac{2nL-y}{\sqrt \varepsilon} \,\,\sigma}\,
\biggr] \, \biggr\},    \]

\vs \vs \no the   formal solution $ \hat u(x,s)$ of  the problem $ ( \hat P_N) $ can be given the form:

\beq     \label{25}
\hat u (x,s) = \,\int _0^L \, [\,\hat \theta_0\,(\,|x-\xi|, \,s\,)\,+\,\,\,\hat \theta_0\,(\,|x+\xi|,\, s\,)\,] \, \,[\,u_0(\,\xi\,) \,+\,\hat f(\,\xi,s)\,]\,d\xi\, -
\eeq
\[  \,  -\,\,\ 2 \,\,\varepsilon \, \,\hat \psi_1 \,(s) \,\, \hat  \theta_0 (x,s)\,+ \, 2 \,\, \varepsilon  \,\, \hat \psi_2 \, (s)\,\,\hat  \theta_0 \,(L-x,s\,).\, \,\,
\]

\vs \section{Explicit solution in the linear case and asymptotic properties}  

\setcounter{equation}{0}
\setcounter{figure}{0}

\setcounter{definition}{0}
\setcounter{notation}{0}
\setcounter{theorem}{0}
\setcounter{esempio}{0}
\setcounter{ex}{0}
\setcounter{proposition}{0}
\setcounter{criterio}{0}

\vs The fundamental solution $ K_0(x,t)  $ of the linear operator  $\, \cal L \,$ defined in (\ref{11}) has been already obtained in \cite{dr8} and it is:

 \beq      \label {31}
 K_0(r,t)=  \frac{1}{2 \sqrt{\pi  \varepsilon } }\biggl[ \frac{ e^{- \frac{r^2 }{4 t}-a\,t}}{\sqrt t}-\,\sqrt b \int^t_0  \frac{e^{- \frac{r^2}{4 y}\,- ay}}{\sqrt{t-y}} \, e^{-\beta (\, t \,-\,y\,)}  J_1 (2 \sqrt{\,by\,(t-y)\,})\,\,dy \biggr],
\eeq

\vs\vs  \no where   $\, r \, = |x| \, / \sqrt \varepsilon \, \, $ 
 and  $ J_n (z) \,$  denotes the Bessel function of first kind.

\vs\vs  \no More, the  following  theorems have been proved in \cite{dr8}:

 \begin{theorem}
In the half-plane $ \Re e  \,s > \,max(\,-\,a ,\,-\beta\,)\,$    the Laplace integral  $\,{\cal L  }_t\,\,K_0 (r,t)\, \,$  converges absolutely for all  $\,r>0,\,$   and it results:

\beq      \label{32}
\,{\cal L  }_t\,\,K_0\,\equiv \,\,\int_ 0^\infty e^{-st} \,\, K_0\,(r,t) \,\,dt \,\,=  \, \frac{e^{- \,r\,\sigma}}{2 \, \sqrt\varepsilon \,\sigma \,  }.
\eeq
\hbox{}\hfill\rule{1.85mm}{2.82mm}
\ethe

 \bthe
 The function  $\  K_0 \, $  has the same  basic properties of the fundamental solution of the heat equation, that is :

i) \hspace{2mm} $ \,\,K_0(x,t) \, \, \in  C ^ {\infty} \,\,\,\,${\em for} \,$\,\,\, t>0, \,\,\,\, x \,\,\, \in \Re. $ 

ii) \hspace{2mm}  For fixed $\, t\,>\,0,\,\,\, K_0 \,$  and its derivatives are vanishing esponentially fast as 

\hspace{8mm}$\, |x| \, $  tends to infinity.

iii) \hspace{2mm}  For any fixed $\, \delta \,>\, 0,\, $  uniformly for all $\, |x| \,\geq \, \delta,\, $it results:

\beq                    \label{33}
 \lim _{t\, \downarrow 0}\,\,K_0x,t)\,=\,0,
\eeq

\vs               
iv) \hspace{2mm} {\em For }$\,t\,>\,0,\,$  {\em it is } $\,\,\, {\cal L} \,K_0   =\, 0.  \,\,$
\hbox{}\hfill\rule{1.85mm}{2.82mm}
\end{theorem}

\vs \vs \no Moreover, if $\,\, \omega = min(a,\beta)\,\,$ and  one puts

\vs
\beq      \label{34} 
E(t) \,=\, \frac{e^{\,-\,\beta t}\,-\,e^{\,-\,at}}{a\,-\,\beta}\,\,>0\,,\qquad \beta _0 =\,\, \frac{1}{a}\, +\, \pi \sqrt b \, \, \displaystyle {\frac{a+\beta}{2(a\beta)^{3/2}}},
\eeq

\vs  \no then  the  following estimates hold \cite{dr8}:

\beq               \label{35}
|K_0| \, \leq \, \frac{e^{- \frac{r^2}{4 t}\,}}{2\,\sqrt{\pi \varepsilon t}} \,\, [ \, e^{\,-\,at}\, +\, b t \,E(t)\, ];  \quad \qquad \int_0^t\,d \tau\, \int_\Re |K_0(x-\xi,t)| \, \,d\xi \leq \,  \beta_0
\eeq

\vs 

\beq               \label{36}
\int_\Re\,\,|\,K_0(x-\xi,t\,)\,|\,\,d\xi\,\,\leq \, e^{\,-\,at}\, +\, \sqrt b\, \pi \,t \,  \, e^{\,-\,\omega \, t }. \,
\eeq

\vs \vs  In order to obtain the inverse formulae for (\ref{25}),          let apply (\ref{32}) to (\ref{24}). Then  one deduces the  following function symilar to  {\em theta functions}:

\beq     \label{37}
\theta_0 (x,t) \,=\,  K_0(x,t) \ +\, \sum_{n=1}^\infty \,\, \ [\, K_0(x \,+2nL,\,t) \, + \, K_0 ( x-2nL, \,t)\,] \, =  
\eeq
\[\, =\sum_{n=-\infty }^\infty \,\, \ K_0(x \,+2nL,\,t). \,\]

 \vs As consequence, by (\ref{25}), the explicit solution of  the {\em linear} problem $\, (P_N )\,$ where $\, F\, =\, f(x,t) \,$ is : 

\beq   \label{38}
 u(\, x,\,t\,)\, = \,\,\int^L_0 \, [\theta_0 \,(|x-\xi|,\, t)\,+ \theta_0 (x+\xi,\,t)\,]\, \,u_0(\xi)\,\, d\xi \,\,+ \,
\eeq

\[ - \,2 \, \varepsilon \,\int^t_0 \theta_0\, (x,\, t-\tau) \,\,\, \psi_1 (\tau )\,\,d\tau\,+\, 2\,\, \varepsilon \int^t_0 \theta_0\, (L-x,\, t-\tau) \,\,\, \psi_2 (\tau )\,\,d\tau\,\]

\[ +\, \,\int^t_0 d\tau\int^L_0 \, [\,\theta_0\, (|x-\xi|,\, t-\tau)+ \theta_0 (x+\xi,\,t-\tau )] \,\,\, f\,(\,\xi,\tau\,)\, \,\,d\xi.\]

\vs\,

\vs Owing to the basic properties of $ K_0(x,t), $ it is easy to deduce the following theorem:

\vs  \bthe
When the linear source $ \, f(x,t)\,  $ is continuous in $ \Omega_T\, $ and the initial boundary  data $ u_0(x),\,\, \psi_i\,(t)\, \,(i=1,2)\, \,   $ are continuous, then  problem $ (P_N) $  admits a unique regular solution $ u(x,t)  $ given by (\ref{38}).
\hbox{}\hfill\rule{1.85mm}{2.82mm}
\ethe

\vs As consequence of the properties of fundamental solution $ K_0 (x,t),$ various estimates for $ u,\,\, u_t,\,\,u_x...  $ could be
obtained. 

As an example, let evaluate  the asymptotic properties of the terms caused by the initial datum $ u_0(x) $  and the source $ f(x,t).\,\, $ If

\[\,\,\,||\,u_0\,|| \,= \displaystyle \sup _{ 0\leq\,x\,\leq \,L\,}\, | \,u_0 \,(\,x\,) \,|, \,\,\qquad||\,f\,|| \,= \displaystyle \sup _{ \Omega_T\,}\, | \,f \,(\,x,\,t\,) \,|, \, \]

\vs \no it results:

\bthe
  When $ \psi_i\,=\,0 \, \,\,\,(i=1,2),\, $   the solution (\ref{38}) of $ (P_N) $, for large $ t , $  verifies  the following estimate:

 \beq   \label{39}
|u(x,t)| \, \leq \,\, 2 \,\,\bigl[\,\,||\,f\,|| \,\, \beta_0\,+\, \,  \,\,||\,u_0\,|| \,\, ( 1\, +\, \sqrt b \,\pi \,t \, ) \,\,e^{- \omega \, t\,}\, \bigr]
\eeq

\vs \no  where $ \, \omega = \min \,(a, \beta )$ and  $ \beta_0\,$
is defined by $(\ref{34})_2$.
\ethe
Proof: Properties of $ K_0(x,t)$ imply that:

\vs 
\beq               \label{310}
\biggr| \,  \int_0^L\,\theta_0 (|x-\xi|,\,t)\,\,d\xi\,\biggl|\,\,\, \leq \,\sum_{n= -\infty }^\infty \, \, \int_0^L\,| K_0(|x-\xi +2nL|, \,t)| \,\,d\xi\,\ =
\eeq

\vs 
\[ =\,\sum_{n= -\infty }^\infty \, \, \int_{x+(2n-1)L}^{x+2nL}\,| K_0(y,\,t)| \,dy\,\,\,\,\leq  \,\, \,\int_\Re\,\,|K_0(y,t)|\,d y .\,\,\]

\vs \vs \no So,  appling properties   $ (\ref{35})_2   $ and (\ref{36})  to  (\ref{38}), the  estimate (\ref{39}) follows. 
\hbox{}\hfill\rule{1.85mm}{2.82mm}

\vs
\section{The Fitzhugh - Nagumo model. A priori estimates }

\setcounter{equation}{0}
\setcounter{figure}{0}

\setcounter{definition}{0}
\setcounter{notation}{0}
\setcounter{theorem}{0}
\setcounter{esempio}{0}
\setcounter{ex}{0}
\setcounter{proposition}{0}
\setcounter{criterio}{0}

\vs Consider now the non linear case of the (FHN) model defined by (\ref{12}). By means of the previous results we are able to obtain integral equations for the two components  $ (u,v) \,$   in terms of the data.  All this implies the qualitative analysis of the solution  together with a priori estimates.

\vs \no At first let us observe that by $(\ref{12})_2 $ one has:

\beq      \label{41}
v\, =\,v_0 \, e^{\,-\,\beta\,t\,} \,+\, b\, \int_0^t\, e^{\,-\,\beta\,(\,t-\tau\,)}\,u(x,\tau) \, d\tau
\eeq

\vs \no and this formula, together with (\ref{14}) require the presence of  the following convolutions:

 \beq     \label {42}
 K_i( r,t) \, = \,\,\int^t_0 \,\,e^{-\,\beta \,(\,t-\tau)\,}\,K_{i-1}\,(x,\tau\,) \, d \tau\,\, \qquad ( i=1,2) 
 \eeq

\vs \vs \no which explicitly are given by \cite{dr8}:

\beq   \label{43}
K_i = \int_0^t\,\frac{e^{- \frac{x^2}{4\varepsilon y}- a\,y\,-\,\beta(t-y)}}{2\,\sqrt{\,\pi\,\varepsilon \,y}} \, \biggl(\sqrt{\frac{t-y}{b\,y}} \biggr)^{i-1}\, J_{i-1 } (2 \sqrt{\,b\,y\,(t-y)\,})\,dy \, \,\,\,\, \,\, (i=1,2).
\eeq

\vs\vs \no As consequence, together with $ \theta_0 $ defined by (\ref{37}), the other two $ \theta  $ functions

\beq  \label{44}
 \theta_i (x,t)\,=\,  \sum_{n=-\infty }^\infty \,\, \ K_i(x \,+2nL,\,t) \, \qquad (i=1,2)
 \eeq

\vs \no must be considered. 

\vs  To allow a plainer reading let's set 
 
\beq      \label{45}
 G_i(x,\xi, t) \, = \, \theta_i\,(\,|x-\xi|, \,t\,)\,+\,\,\,\theta _i \,(\,x+\xi,\,t\,)\,\qquad (i=0,1,2)
\eeq

\vs \vs \no In this manner, owing to  (\ref{38}) one has:

\beq   \label{46}
 u(\, x,\,t\,)\, = \,\,\int^L_0 \, [\,G_0(x,\xi,\, t)\,\, \,u_0(\xi)\,\,- \,G_1(x,\xi,\, t)\, \,v_0(\xi)\,\,] d\xi \,+\,\eeq
\[ - \,2 \, \varepsilon \,\int^t_0 \theta_0\, (x,\, t-\tau) \,\,\, \psi_1 (\tau )\,\,d\tau\,+\, 2\,\, \varepsilon \int^t_0 \theta_0\, (L-x,\, t-\tau) \,\,\, \psi_2 (\tau )\,\,d\tau\,\]

\[ +\, \,\int^t_0 d\tau\int^L_0 \, \,G_0(x,\xi,\, t-\tau) \,\,\,  \varphi\,[\,\xi,\tau,\,u(\xi, \tau)]\,  \,\,d\xi\,.\,\]

\vs\,\vs  As for the  $ v $ component, by  (\ref{41}) one deduces:

 \beq      \label{47}
v(x,t) \, =\,\, v_0 \, e^{\,-\,\beta\,t\,}\,\,+\, b\, \,\,\int^L_0 \, [G _1\,(x,\xi,\, t)\,\, \,u_0(\xi)\,\,- G _2\,(x,\xi,\, t)\,\,  v_0(\xi)\,] d\xi \,\,+ \,
\eeq 
\[ - \,2 \,b\,\, \varepsilon \,\int^t_0 \theta_1\, (x,\, t-\tau) \,\,\, \psi_1 (\tau )\,\,d\tau\,+\, 2\,\,b\,\, \varepsilon \int^t_0 \theta_1\, (L-x,\, t-\tau) \,\,\, \psi_2 (\tau )\,\,d\tau\,\]

\vs
\[ +\,b  \,\int^t_0 d\tau\int^L_0 \, \,G_1\, (x,\xi,\, t-\tau)\,\,\, \varphi\,[\,\xi,\tau,\, u(\xi, \,\tau)\,]\,d\xi.\,\, \]

\vs\vs\vs  Let us observe that the kernels $ K_1(x,t)  $ and $ K_2(x,t)  $ have the same  properties of $ K_0(x,t).  $ In fact \cite{dr8}:

\vs \bthe

For all the positive constants $ a, \,b\, \varepsilon, \beta $  it results:

\beq   \label{48}
\int_\Re |K_1| \, \ d\xi \leq \, \,E(t);\,  \qquad  \int _0^t\\d\tau \,\int_\Re |K_1| \, \ d\xi \leq \, \beta_1\,
\eeq

\beq   \label{49}
\int_\Re \left|K_2 (x-\xi,t)\right| \, d\xi \, \leq \, \int_0^t \, \, e^{\,-\,ay\, - \beta(t-y)\,}\,(\,t-y\,) \,dy \,\leq \, t\, E(t)
\eeq

\vs \vs \no where $ E(t) $ is defined in $(\ref{34})_1$ and $ \beta_1\,=\, ({a\,\beta})^{\,-1}.\,$ \hbox{}\hfill\rule{1.85mm}{2.82mm}
\ethe

\vs\vs  Now, let $\,\, ||\,z\,|| \,= \displaystyle \sup _{ \Omega_T\,}\, | \, z \,(\,x,\,t) \,|, \,\, $ and let $ \,{\cal B}_ T \,  $ denote the Banach space 

\beq   \label{35}
  \,{\cal B}_ T \, \equiv \, \{\, z\,(\,x,t\,) : \, z\, \in  C \,(\Omega_T),  \, \,\,   ||z|| \, < \infty \ \}.
\eeq

\vs\vs   By means of standard methods related to integral equations  and owing to basic properties of $ K_i,\,\, G_i \,\, (i=0,1,2)$  and $ \varphi(u),\,\, $it is easy  to prove that the  mapping defined by (\ref{46}) ia a contraction  of $ {\cal B}_ T $ in $ {\cal B}_ T  $ and so it admits an unique fixed point   $ u(x,t)  \, \in {\cal B}_ T $  \cite{c,dmm}.  Hence

\vs
\bthe
When the initial data $ (u_0, v_0 )$ are continuous functions, then the Neumann problem related to the non linear (FHN) system (\ref{12}),(\ref{13}) has a unique solution in the space of  solutions which are regular in $ \Omega_T $. 
 \hbox{}\hfill\rule{1.85mm}{2.82mm}
 \ethe

\vs \vs Continuous dependence for the solution of  $ (P_N) $ is an obvious consequence of the previous estimates. As an example  of asymptotic properties let us consider the case  $ \psi_1\,=\,\psi_2\,=\,0\, $ and let

\[
||\,\varphi\,|| \,= \displaystyle \sup _{ \Omega_T\,}\, | \,\varphi \,(\,x,\,t,\,u) \,|, \, \]

\vs \no then by means  of (\ref{46}), (\ref{47}) and  owing to the estimates (\ref{35}), (\ref{36}), (\ref{48}), (\ref{49}), the following theorem can be stated:

\bthe
For regular solution $ (u,v) $ of the (FHN) model, when  $\psi_1\,\,= \psi_2 \,= \,0,\, \, $  the following estimates hold:

\vs 
\beq            \label{510}
\left\{ 
 \begin{array}{lll}                                                   
 \left| u \, \right| \, \leq  2\,[\,\left\| u_0 \right\| \, (1+\pi \sqrt b \, t ) \, e^ {\,-\omega\,t\,}\,+\,\left\| v_0 \right\|\,E(t) \, +\, \beta_0 \,\left\| \varphi \right\|\,] 
   \\
\\
\left| v \, \right| \, \leq  \left\| v_0 \right\|\, e^ {\,-\,\beta\,t\,}\,+\,2\,[\,b\,(\,\left\| u_0 \right\|\,+\, t\, \left\| v_0 \right\|\,) \, E(t) \, + \, b\, \beta_1\, \left\| \varphi \right\| \,]
\\ 
   \end{array}
  \right.
 \eeq
\hbox{}\hfill\rule{1.85mm}{2.82mm}
\ethe

\vs \no Therefore, when $ t $ is large, {\em the effect due to the initial disturbances $\, (\,u_0, v_0\,) \, $ is exponentially  vanishing  while   the effect of the non linear source is bounded for all $ t. $}

\vs All   the previous results  can be applied to the boundary  Dirichelet or mixed conditions, too.

\vs \vs\no 
{\bf Acknowledgements}
This work has been performed under the auspices of the G.N.F.M. of   I.N.D.A.M. and M.I.U.R. (P.R.I.N. 2009) "Waves and stability in continuous media".

\vs \no The author thanks Professor P. Renno for his useful suggestions.

\end{document}